\documentclass[namedreferences]{solarphysics}
\usepackage[optionalrh,natbib,solaromanenum]{spr-sola-addons} % For Solar Physics
\usepackage[pdfborder={0 0 0 },urlcolor=blue,breaklinks]{hyperref}
\usepackage{graphicx}        % For eps figures, newer & more powerfull
\usepackage{epsfig}          % For eps figures, old commands
\usepackage{color}           % For color text: \color command
\usepackage{url}             % For breaking URLs easily trough lines
\newcommand{\rsun}{{\rm R}_\odot}

\newcommand{\eg}{{\it e.g.}}
\newcommand{\etc}{{\it etc}}

\begin{document}

\begin{article}

\begin{opening}

\title{The Quest to Understand Supergranulation and Large-Scale Convection in the Sun}

\author{Shravan~M.~\surname{Hanasoge}$^{1,2,3}$\sep
Katepalli~R.~\surname{Sreenivasan}$^{4}$
}
\runningauthor{S.~M.~Hanasoge, K.~R.~Sreenivasan}
\runningtitle{Supergranulation and Global-Scale Convection in the Sun}

\institute{$^{1}$Tata Institute of Fundamental Research, Mumbai, India\\
	      $^{2}$Max-Planck Institute for Solar System Research, G\"{o}ttingen, Germany\\
	      $^{3}$Department of Geosciences, Princeton University, NJ, USA\\
                     email: \url{hanasoge@tifr.res.in}\\
                 $^{4}$ Departments of Physics and Mechanical Engineering and the Courant Institute of Mathematical Sciences, New York University, NY, USA\\
                 $\bigstar$ Invited Review }

\begin{abstract}
Surface granulation of the Sun is primarily a consequence of thermal
transport in the outer 1\,\% of the radius. Its typical scale of
about 1 -- 2\,Mm is set by the balance between convection, free-streaming
radiation, and the strong density stratification in the surface layers.

The physics of granulation is well
understood, as demonstrated by the close agreement between numerical
simulation, theory, and observation. Superimposed on the energetic
granular structure comprising high-speed flows, are larger
scale {long-lived flow systems ($\approx 300\,{\rm m s^{-1}}$)} called supergranules. Supergranulation has a typical
scale of 24\,--\,36\,Mm. It is not clear if supergranulation results from the
interaction of granules or is causally {linked} to deep
convection or a consequence of magneto--convection. Other outstanding questions
remain: how deep are supergranules? How do they
participate in global dynamics of the Sun?
Further challenges are posed by our lack of insight into the dynamics of larger scales in the deep convection region. Recent helioseismic constraints have
suggested that convective
velocity amplitudes on large scales may be overestimated by an order of magnitude or more, {implying} that
Reynolds stresses associated with
large-scale convection, thought to play a significant role in the sustenance
of differential rotation and meridional circulation, might be two orders of magnitude weaker than {theory and computation predict}.
{While basic understanding on
the nature of convection on global scales and the maintenance of
global circulations is incomplete, progress is imminent, given substantial improvements in computation, theory and
helioseismic inferences.}
\end{abstract}
\end{opening}

\section{Introduction}
     \label{intro}
The transport of thermal energy from the solar interior to its surface drives, directly and otherwise, a broad range of dynamical processes in the Sun. From the core outwards, transport is achieved primarily by radiation: photons diffusing by free--free radiative scattering. Plasma in this region is almost fully ionized. In the outer parts of the radiative region, at distances of roughly 500 Mm from the center, the mean temperature is low enough that the opacity locally increases due to bound--free radiative absorption of photons by heavy elements. This process gradually diminishes the effectiveness of radiative thermal transport, resulting in the onset of convection. The convection zone, comprising the outer 30\,\% of the Sun's radius, is an optically thick medium where heat transport is accomplished by convective motions of plasma.
	
The physics of convection in the {low-viscosity {large} density- and temperature-gradient regime} associated with the convection zone is not well known. Most of the convection zone is moderately stratified but, as plasma approaches the photosphere, it undergoes rapid expansion due to {steep} near-surface density gradients. Helium- and hydrogen-ionization zones, which form amid the cooling plasma in the near-surface layers, are thought to power various scales of turbulent convection. Eventually, across a thin shell of 300 km thickness, the thermal-transport mechanism transitions from convection to free-streaming radiation in the optically thin photosphere. The interplay between the equation of state (that locally determines opacity) and thermal and hydrostatic stratification is critical to controlling the physics of granulation. A detailed review on the properties of granules has been given by, {\eg}, \citet{nordlund09}.
	
The observed properties of energy-carrying {granules}, such as their distribution in size, radiative intensity, and spectral-line formation are accurately reproduced by numerical simulations \citep[{\eg}][]{stein00,voegler2005}. This appears to be the case despite the fact that the fluid is highly stratified, exhibits complicated energetics, and is subject to non-standard boundary conditions (stress-free or no-slip constitute examples of ``standard" boundary conditions). The success of simulations is likely due to two aspects of physics that are correctly incorporated: {a complicated equation-of-state in the very thin upper thermal boundary layer (0.3 Mm; compare with the radius of the Sun, $\rsun \approx 700$ Mm) and sharp stratification arising from the short near-surface density scale height ($\approx$0.15 Mm)}. The former regulates convective--radiative transport balance while the latter constrains near-surface fluid dynamics. In particular, the large density contrast also appears to block strong interactions between scales, leading to a well structured photospheric flow field. It appears that incorporating the ingredients of an accurate equation of state and background stratification leads to a high-fidelity reproduction of line formation and spatial scales.	
	
A spatio--temporal power spectrum of photospheric flows \citep[\eg][]{hathaway00} reveals, in addition to granular scales, a distinct peak at much larger scales, termed supergranulation. Supergranular physics is deserving of a detailed exposition all to itself \citep[\eg][]{supergranulation10}. It has evaded easy characterization as a number of fundamental questions on supergranulation remain {unanswered}: for example, do magnetic fields regulate their dynamics; are they driven by mechanisms similar to those of the granular structure; are they related to the scales of deep convection. To answer the last question, one must understand how to model convection in the solar interior. {This} presents a formidable challenge. Unlike  the situation in the near-surface layers, it is not apparent that there exist dominant physical ingredients that are easily modeled to yield the right results. The stability of descending plumes is likely to be an important factor \citep[][]{spruit97, rast98}, but how to model this in the extreme parametric regime of solar convection is unclear.	

This review attempts to summarize the state of the art in observations and theory of intermediate-to-large scale flow systems in the Sun. However, it should be noted that there have been a number of reviews, in the past two decades, on convection and supergranulation in the Sun. Consequently, this review article does not aim to overly emphasize the topics of photospheric observations and the detailed physics of convection. Instead, we will briefly survey what has been observed, theorized, and numerically modeled of convection and discuss the possibilities of how helioseismology can shed light on these topics. Seismic measurements, if interpreted correctly, can provide substantial information about properties of the interior \citep[\eg][]{jcd02, gizon05,gizon2010}.

\subsection{Turbulence}

Since we will often use the term ``turbulence" in this article, it is worth exploring its meaning briefly. Turbulence, as understood by classical fluid mechanicians, refers to a state of flow in which interactions within a large range of spatial and temporal scales are important to determining dynamics. The classical Kolmogorov picture of fluid turbulence consists of the energy being input at the largest scales (of size comparable to the entire system) and of this energy being transferred sequentially to smaller scales until molecular viscous dissipation becomes important \citep[\eg][]{frisch95}. This range of scales is captured by the relation
	\begin{equation}
	\frac{L_{\rm D}}{L} = Re^{-\frac{3}{4}},~~~~~~~~~Re = \frac{u L}{\nu}\label{reynolds}
	\end{equation}
where $L_{\rm D}$ is the dissipation scale, $L$ is the scale of the system itself, $u$ is a characteristic velocity, $\nu$ is the viscosity, and $Re$ is the Reynolds number, a non-dimensional quantity that characterizes the degree of turbulence. {Equation~(\ref{reynolds}) states that as the Reynolds number increases, the dynamical range of scales also grows}. {But we should remember that Kolmogorov phenomenology does not apply to all turbulent flows.}
	
Since the theory of these strongly interacting scales is not yet on a firm footing, practitioners resort to laboratory measurements (where Reynolds numbers are low to moderate) and to direct numerical simulations (also with low to moderate Reynolds numbers) that aim to mimic the interaction among the full dynamical range of scales. Faithful replication of this feature is not possible today even on the world's most powerful computers for astrophysical Reynolds numbers. Even for most industrial applications, where Reynolds numbers are substantially smaller than in astrophysics, simulations must make a number of approximations. In particular, simulating convection in the Sun over the full dynamical range is not feasible in the foreseeable future.

Nonlinear simulations of convection and dynamo action in the Sun are capable of capturing only a small range of the largest scales of the system under study.
The hope is that modeling this finite scale range is sufficient to understand the large-scale dynamics of the system. This proves successful in the case of granulation, as discussed in the introductory section, but it is unclear whether global solar phenomena are also controlled by similarly spatially band-limited processes.
{Consider the scenario, for instance, where the global circulations of differential rotation and meridional circulation are forced by Reynolds stresses over a
very broad range of scales, extending from the small (granulation) to the large \citep[an idea explored by, \eg][]{miesch12}. Numerical calculations that only model the largest scales, as opposed to those that include all of the relevant scales (both large and small), would each
predict different dynamics. Thus it is important to identify the relevant set of ``turbulent" scales for the problem so that the outcome contains the correct physics.}

\section{Supergranulation}
\subsection{Observational Evidence}
\begin{figure}
   \centerline{
               \includegraphics[width=0.99\textwidth,clip=]{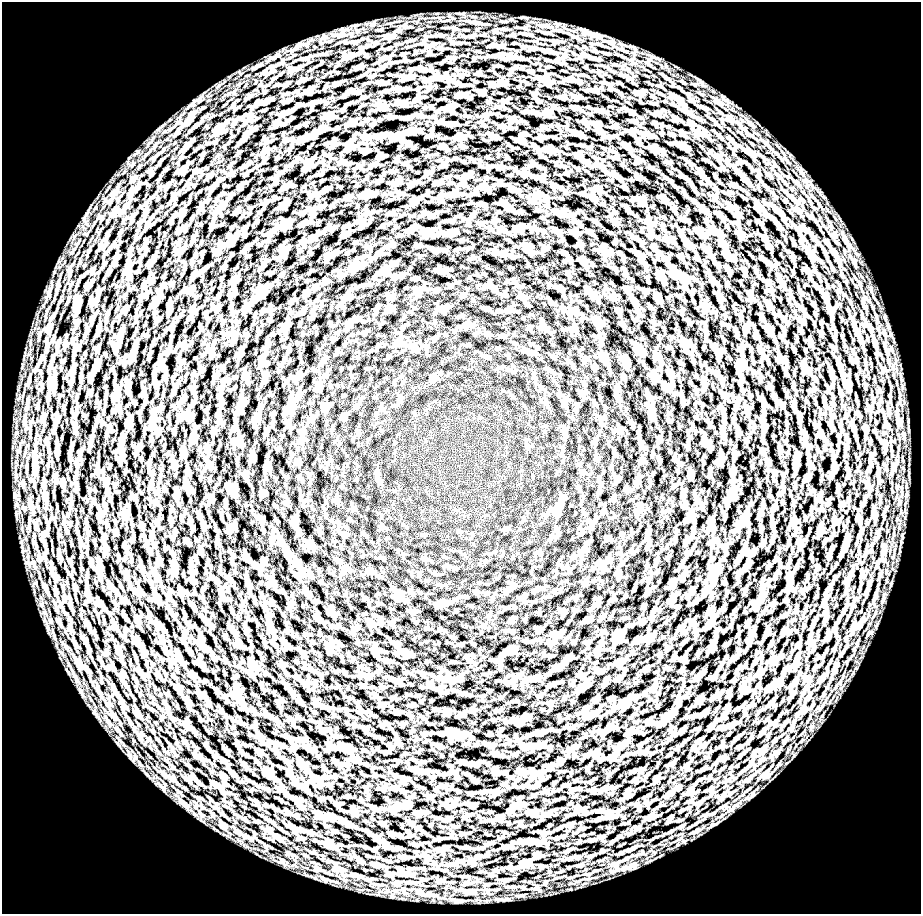}}\vspace{0.5cm}
   \caption{Line-of-sight Dopplergram taken by the {\it Michelson Doppler Imager} onboard the {\it Solar and Heliospheric Observatory}.
   		The greyscale spans the range $\pm350\,{\rm ms^{-1}}$.
   		Average solar rotation has been subtracted; it is seen that supergranules cover the photosphere. Because vertical
		velocities associated with supergranules are around 4 ${\rm ms^{-1}}$ and horizontal velocities are about 300 ${\rm ms^{-1}}$ (see Table~\ref{superg.table}), their visibility is
		significantly reduced at disk center. Figure courtesy of M. Rieutord, also see \citet{rieutord10}.}\label{supergranulation}
  \vspace{-0.2cm}
  \end{figure}

Supergranulation, shown in Figure~\ref{supergranulation}, is such a highly visible feature that it was observed early in the previous century by \citet{plaskett16}.
It was (re)discovered by \citet{hart56} and characterized further by \citet{leighton62}. See the article by \citet{supergranulation10} for a detailed review on the decades of
observational and theoretical research directed at understanding supergranulation.

\begin{figure}
   \centerline{
               \includegraphics[width=0.99\textwidth,clip=]{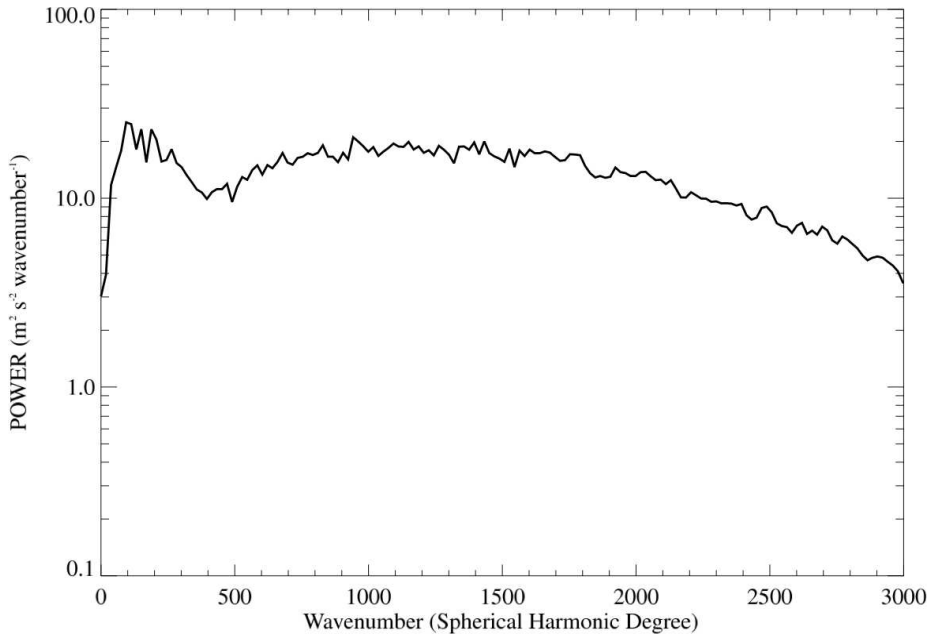}}\vspace{0.5cm}
   \caption{Power spectrum of photospheric convection, estimated using Dopplergrams taken by the {\it Michelson Doppler Imager}
   onboard the {\it Solar and Heliospheric Observatory} \citep[courtesy, M. Rieutord;][]{hathaway00}. A peak associated with supergranulation is seen at $\ell \approx 120$ but
   power falls linearly at larger scales, i.e. as $\ell \rightarrow 0$. The interpretation of this spectrum is complicated by
   that fact that the {\it Michelson Doppler Imager} takes line-of-sight measurements at the photosphere,
   resulting in different components of the flows, i.e. horizontal or radial, being recorded
   at different positions on the solar disk.}\label{spectrum}
  \vspace{-0.2cm}
  \end{figure}

Supergranulation is a stochastic process in the variables of emergence, location, and size \citep[\eg][]{schrijver97}, so measurements of properties pertaining to radiative intensity, flow magnitude, and size of individual supergranules are intrinsically dependent on a number of details such as the methods applied, the degree of smoothing, averaging, \etc. What makes the study of supergranulation possible is statistical averaging over measurements of large numbers of individuals. This averaging, when applied to vast databases of observations taken by instruments such as the {\it Michelson Doppler Imager} \citep[MDI:][]{scherrer95} and the {\it Helioseismic and Magnetic Imager} \citep[HMI:][]{hmi} over the past 17 years, allows for inferences with high statistical significance. {Variations in the properties of supergranulation as a function of the solar cycle can also be measured \citep{berrilli99}.} Figure~\ref{spectrum} shows the power spectrum of photospheric convection measured by the MDI. A clear peak associated with supergranules is seen at $\ell \approx 120$ but power in larger scales is weak, with the spectrum falling linearly towards $\ell = 0$. Various authors characterize the statistical distributions of area and lifetime of supergranules \citep[\eg][]{schrijver97, hagenaar97, berrilli99, berilli04, gizon08, meunier08}. The size of a typical supergranule is roughly 30 Mm. Table~\ref{superg.table} summarizes the known, observed properties of supergranules.

	\begin{table}
	\caption{ Photospheric properties of supergranulation. Several different
	methods are applied to obtain these measurements. For sizes, see, \eg\, \citet{schrijver97, gizon08, rieutord10},
	lifetimes: \citet{gizon06_esa, gizon08}, horizontal velocity: \citet{rieutord10}, vertical velocity: \citet{duvall10} and
	intensity variation: \citet{rast09}.
	}
	\label{superg.table}
	\begin{tabular}{ccclc}     
                           
	  \hline                   
	Scale & Horizontal Velocity & Vertical Velocity & Lifetime & Intensity variation \\
	       Mm     & [ m/s ] & [ m/s ] & [ hours ] &   [ K ] \\
	  \hline
	24\,--\,36 & 300 &
	4 & 24\,--\,96  & 1\tabnote{Ambiguities still persist; see Goldbaum et al. (2009)} \\
	  \hline
	\end{tabular}
	\end{table}

Supergranules are regular flow systems comprising a central region of upflow ({where ``up" means from the interior to the photosphere}),
horizontally diverging flows, and downflows at the edges.
Small magnetic elements occupy downflow lanes between adjacent supergranules, enhancing the visibility of the boundaries.
{Measuring the intensity contrast of a supergranule is complicated by the accumulation of magnetic flux at the boundaries. Small flux concentrations tend to 
locally reduce plasma density, lowering the opacity and allowing radiation to escape from layers deeper than the photosphere (and therefore hotter).
A na\"{i}ve observational analysis would therefore show that the edges of the supergranule are ``hotter" than the center.
\citet{rast09}, analyzing some $10^5$ supergranules and applying statistical corrections to remove the ``hot" boundaries,
found that the center is slightly warmer than the edge by about 1 K (in a purely hydrodynamic sense).
However, as \citet{rast09} note, this intensity contrast is so small that one cannot unambiguously conclude that supergranulation is indeed thermally driven.
The lack of an obvious thermal signature and the fact that downflows are co-spatial with magnetic elements
have led many to believe that supergranulation is a manifestly magneto--convective phenomenon \citep[\eg][]{berrilli99}.}
	
\subsection{Theories of Supergranulation}

A successful theory of supergranulation must rely on weak thermal driving and produce a characteristic size of about 30 Mm with time and velocity scales noted in Table 1. If supergranulation is a purely hydrodynamic phenomenon, then the photospheric magnetic network is broken up by the flows. However, as mentioned above, a number of theories consider {supergranulation} to be magneto--convection, in which case magnetic tubes that cluster along downflow lanes play an {important} role in regulating supergranulation physics.
	
\citet{rast03} posited that granulation-related downflow plumes merge at depth, resulting in larger, long-lived downflow regions, in turn leading to the formation of supergranular flow systems. Based on a kinematic formulation of this theory, \citet{rast03} was able to recover the scales of meso- and supergranulation. The horizontal divergence associated with these flow systems would then sweep magnetic elements into the boundaries, enhancing the visibility of supergranules. {Thus, \citet{rast03} has suggested that supergranulation is a hydrodynamically driven  phenomenon.}
	
Supergranular size varies with the solar cycle \citep{berrilli99}, decreasing during solar maxima, leading to the
possibility that magnetic fields dynamically alter supergranular properties {(however, plage area increasing during solar maximum could be the source of the observed corresponding decrease in supergranular size, i.e. it is the increase in plage area that we are effectively recording, not diminished supergranules [M. Sch\"{u}ssler 2013, {private communication}]). To explain this variation, \citet{crouch07} put forth an interaction theory where supergranular flows are caused by the clustering of small-scale magnetic elements.}
	
Numerical simulations of surface convection show preliminary signs of supergranular flows \citep[\eg][]{stein12} although horizontal computational domain sizes are just getting large enough to accommodate a few supergranules. Perhaps in the future, with much bigger and more highly resolved computational domains, we may be able to replicate supergranulation.
{Simulations are not in agreement on the causal mechanism for supergranulation: magneto--convection \citep[{private communication}, T. Yokoyama 2013; also][]{stein12} {\it versus} helium ionization ({private communication}, M. Rempel 2013).}
Helioseismology has the (potential) discriminating power to eliminate or add weight to specific theories because it can ostensibly be used to infer average depth, magnetic field and thermal conditions within a supergranule. If we are successful at determining these properties, then a number of theories can be put to test.
	
\section{Helioseismology}

A number of helioseismic investigations of supergranulation may be found in the literature \citep[\eg][]{Kosovichev1997, duvall00,gizon03,zhao03,braun04, gizon08,duvall10,svanda12,duvall12,dombroski13}. Most of these post-date the SOHO and SDO missions and a large fraction of them perform statistical studies of supergranulation, i.e. averaging the seismic analyses of large numbers of individual supergranules. This allows for the measurement of travel-time shifts around the average supergranule at very high signal-to-noise ratios.
	
Using the nearly decade-long repository of MDI observations, \citet{duvall10} analyzed thousands of supergranules and obtained precise measurements of travel times, which were used to infer interior flow. Based on this they were able to constrain the vertical flow at the center of the supergranule to about 4 ${\rm ms^{-1}}$.
Subsequently, based on extensive analyses of travel-time measurements, \citet{duvall12} suggested that supergranular flows peak at a depth of 2\,--\,3 Mm, reaching 600\,--\,800 ${\rm ms^{-1}}$ in speed. \citet{svanda12} reconfirmed these results whereas {\citet{woodard07} disagreed with them.
The analysis of \citet{woodard07} relies on a different seismic measurement, but one that is interpreted carefully, using
wave physics and modeling instrumental systematics. \citet{woodard07} found that supergranular flows decrease in magnitude with depth
and saw no evidence for a return flow within the first 5\,--\,7 Mm of the photosphere. Other results, such as the wave-like properties of supergranulation \citep{gizon03}, are not well understood and, additionally, have proven controversial \citep{lisle04}.}

It is important to consider why there are disagreements. If we were to model systematical and finite-wavelength effects, and inversion nonlinearity, the trustworthiness and accuracy of seismic inferences will greatly improve. While we have a solid understanding of instrumental effects, {finite-wavelength effects are still not routinely modeled. Inversion nonlinearity is generally ignored}. Only recently have theoretical advances \citep{hanasoge11} made it possible to address this significant issue.

\subsection{Finite Wavelengths}
In most practical scenarios involving the propagation of light, its wavelength is small compared to the sizes of objects it propagates through and around. Consequently the propagation of light is typically treated in the infinite frequency (zero-wavelength) limit, termed ray theory or, more formally, the WKBJ limit \citep[\eg][]{gough07}. However, waves that propagate in the Sun have finite wavelengths, comparable in size to the objects that they encounter \citep[\eg][]{birch00, birch01, gizon06_esa}. Despite this, ray theory is widely invoked to interpret travel-time measurements, leading to potentially erroneous inferences. Finite-frequency sensitivity kernels that capture these effects may be computed \citep[\eg][]{dahlen99, gizon02, birch07} but they have not been widely adopted due to the complexity and difficulty encountered in implementing such inversions.
	
\subsection{Nonlinear Inverse Theory}
Helioseismology deals with the solution of high-dimensional inverse problems, where seismic measurables, obtained at the photosphere, must be related to a large number of parameters, \eg\, the sound speed, flows, and magnetic fields in the interior.
Typically, one defines a cost function, such as the $L_2$ norm of the difference between predicted and observed wave travel times, i.e.
\begin{equation}
	\chi = \frac{1}{2}\sum_{i} \tau_i^2,
	\end{equation}
where $\tau_i$ is the misfit in travel times {measured between various point-pairs $[i]$ at the photosphere}. A typical starting model of the Sun {for instance} could be model S \citep[\eg][]{jcd}. The question is how one must change the model of the Sun so as to reduce the misfit functional $\chi$.
	
The classical helioseismic inverse problem is stated as
	\begin{equation}
	\int_\odot {\rm d}{\bf x}\,{\bf K}({\bf x})\cdot {\bf v}({\bf x}) = \{\tau_i\}, \label{smalldev}
	\end{equation}
where the volume of the interior of the Sun is denoted by $\odot$, ${\bf x}$ is the spatial coordinate, ${\bf K}$ is a vector {\it sensitivity kernel}, denoting the finite-wavelength sensitivity of a travel-time measurement $[\tau_i]$, say, to the vector flow $[{\bf v}]$. Equation~(\ref{smalldev}) only applies to small perturbations $[{\bf v}]$---where, by small, we mean that the first-Born, single-scattering (weak) approximation applies to the flow system $[{\bf v}]$. {In contemporary helioseismic literature}, ${\bf K}$ is computed around the quiet Sun (with no flows) and only small deviations $[{\bf v}]$ are considered. However, the Sun can be a strong scatterer, as in the case of sunspots. Many perturbations such as supergranules, typically considered to be weak scatterers, are possibly explained only by considering a strong-scattering approximation. The correct form of Equation~(\ref{smalldev}) should then be
	\begin{equation}
	\int_\odot {\rm d}{\bf x}\,{\bf K}({\bf x}; {\bf v})\cdot {\bf v}({\bf x}) = \{\tau_i\}. \label{largedev}
	\end{equation}
	
A simple analogy to consider is Newton's method applied to determine the roots of a polynomial $f(x) = 0$. This can be a nonlinear inverse problem depending on the nature of $f(x)$. Newton's method to determine the root is straightforward, requiring one to iterate over a sequence of $x_i$, such that
	\begin{equation}
	x_{n+1} = x_n - \frac{f(x_n)}{f'(x_n)},\label{newton}
	\end{equation}
where $f'(x_n)$ is the local gradient, until $f(x_k)$ satisfies a prescribed convergence criterion for some $k$. The local gradient $[f'(x_n)]$ is analogous to the sensitivity kernel in Equation~(\ref{largedev}), i.e. ${\bf K({\bf x}, {\bf v})}$ as is the local solution $x_n$ to ${\bf v}$. Current helioseismic inverse theory allows for taking only one step along the gradient direction. {Further, because the gradient ${\bf K}$ is computed around the quiet Sun background, a single-step inversion is unlikely to accurately
retrieve the subsurface structure of large perturbations such as sunspots and possibly supergranules}. Lastly, an ideal inverse problem should attempt to simultaneously invert for sound speed, flows, magnetic fields, and density \citep{hanasoge12_mag}.
	
Although present practice falls short of this ideal, we may soon be performing such inversions. Terrestrial seismologists routinely invert for the structure of strong heterogeneities in Earth. This is accomplished by performing iterative nonlinear inversions based on the adjoint method \citep[\eg][]{tarantola84, tarantola87, tromp05, CarlTape09}. The case for using the adjoint method is strong, as demonstrated by the success that terrestrial seismologists have had in fitting their data and obtaining consistent results.		
Recently, \citet{hanasoge11} extended the adjoint method to the helioseismic scenario. {While substantial work remains to be done in testing, verifying, and assessing the capabilities of these extensions, the technique holds considerable promise.}
	
\citet{birch_hanasoge_07} analyzed the travel times associated with the average over a large sample of supergranules. Because of substantial averaging, the signal-to-noise ratio of the travel times was very high. Despite using a combination of semi-analytical \citep{birch} and computational methods \citep{Hanasoge_couvidat_2008}, \citet{birch_hanasoge_07} were unable to fit the measurements to an appropriate level of accuracy (A. Birch 2012, {{private communication}}). One source of the discrepancy is that the travel times are nonlinear functions of the $\approx 300\,{\rm ms^{-1}}$ supergranular flow velocities; if so, the adjoint methodology developed by \citet{hanasoge11} may help us in constructing better models of supergranules.
	
\section{Large-Scale Convection and the Sustenance of Global Differential Rotation}

{One of the most important results of helioseismology} is the inference of the internal differential rotation of the Sun \citep{schou98} through studies of global solar oscillations. {Within the convection zone, the rotation rate is seen to be constant on approximately radially directed contours, and not (Taylor--Proudman) rotation-axis-aligned contours. While various mechanisms to explain the appearance of these peculiar rotation contours have been elucidated, none has succeeded in explaining all of the relevant observations}. Simulations, advanced as they may be, are unable to reproduce the full differential-rotation profile, although they are being continuously improved.

The Reynolds-stress tensor, centrally involved in thermal and momentum transport, is given by
	\begin{equation}
	\{{\bf R}\}_{ij} = \rho\langle v'_i v'_j \rangle,
	\end{equation}
where the angular brackets indicate a suitable form of statistical averaging (spatial, temporal, ensemble, \etc.), $v'_i$ are velocity fluctuations along component $i$ around the mean flow state and ${\bf R}$ is the Reynolds-stress tensor. The divergence of the Reynolds stress {constitutes} the turbulent momentum-transport {term}.

\subsection{A Latitudinal Temperature Gradient Must Exist}\label{latgrad}

The Rossby number $[Ro]$, comparing rotation and advective transport in a rotating fluid system, is defined as
	\begin{equation}
	Ro = \frac{U}{\Omega\,L},~~~~{\rm and~~alternatively,}~~~~Ro = \frac{\tau_{\rm rot}}{\tau_{\rm con}},
	\end{equation}
where $U$ is a characteristic fluid (advective) velocity, $L$ the corresponding length scale, and $\Omega$ is the rotation rate. In a convecting system such as the Sun, it may also be defined as the ratio of the rotation timescale [$\tau_{\rm rot}$] to the convective lifetime [$\tau_{\rm con}$]. The mechanics at high Rossby numbers is described by pure fluid effects and knowing how to model ${\bf R}$ becomes critical. At low Rossby numbers, rotation is very important and Reynolds stresses may be ignored. In this scenario, likely applicable to the Sun, the thermal-wind-balance equation effectively describes the internal rotation
	\begin{equation}
	\frac{\partial\Omega^2}{\partial z} = \frac{c}{r^2\sin\theta}\frac{\partial\langle S\rangle}{\partial\theta},\label{TWB}
	\end{equation}
where $z$ is the coordinate along the rotation axis, $r$ is spherical radius, $\theta$ is the co-latitude, $\langle S\rangle$ is the temporally and longitudinally averaged entropy, and $c$ is some constant. Thermal-wind balance can be applied \citep[\eg][]{balbus09} to {describe the observed conical rotation contours, the outcome being that there must be a latitudinal temperature gradient} ({baroclinic forcing}). Since the internal rotation of the Sun [$\Omega$] is helioseismically measured, one can substitute this quantity into equation~(\ref{TWB}) to obtain the required $\langle S \rangle$. Solar-like differential rotation requires having warm poles relative to the Equator, on the order of a few Kelvin \citep{miesch05}. \citet{rast08} measured a 2.5 K photospheric enhancement in the temperature at the poles relative to the Equator and, despite instrumental systematics, concluded that this represented a real temperature difference.	

\subsection{Reynolds Stresses Must Also Participate}\label{stress}

The Sun exhibits a persistent poleward meridional circulation \citep[first measured at the photosphere by][]{Duvall1979}, complicating the picture of Section~\ref{latgrad}. Indeed, \citet{miesch11} and \citet{miesch12}, based on models of the interplay between a latitudinal entropy gradient, Reynolds stress, and meridional circulation, argue that a thermal wind cannot solely set up and maintain differential rotation. Because the Sun is largely a solid-body rotator, the angular momentum at the Equator is larger than that at the poles (larger lever arm: angular momentum $= \rho\,\Omega\, r^2\sin^2\theta$, $\theta =\pi/2$ at the Equator). Poleward meridional circulation, which carries parcels of high angular momentum from the Equator to the low-angular-momentum polar regions, tends therefore to spin up the latter. This would result in anti-solar rotation, with rapidly rotating poles. Further a relatively warm pole, as suggested by equation~(\ref{TWB}) would set up an anti-solar meridional circulation, i.e. toward the Equator at the photosphere. We note here that this picture does not fully reflect the physics of the momentum balance; \citet{miesch11} show that {\it any} meridional circulation, i.e. Equator- or poleward, would drive rotational shear to assume an anti-solar form. The arguments are subtle, and we refer the interested reader to \citet{miesch11} and \citet{miesch12} for further details.
	
Therefore the only way to set up solar-like rotation, argue \citet{miesch12}, is for the Reynolds stresses to play a consequential role in transporting angular momentum. In the schema of the Babcock--Leighton flux-transport dynamo, a widely accepted model of global magnetic field generation in the Sun \citep[\eg][]{dikpati99}, the global polar field is {wound} up by differential rotation at the base of the convection zone to create toroidal fields for the succeeding magnetic cycle. If this is indeed the mechanism by which magnetic fields are regenerated in the Sun \citep[\eg][]{dikpati06}, these fields, acting much like an elastic rubber band, will diminish rotational shear. Reynolds stresses must then power the rotational shear by supplying fresh angular momentum to the Equatorial regions. {It is thus very important to measure properties of convection in the solar interior in order to appreciate whether the kinetic energy and Reynolds stress tensor are of sufficient magnitude to power differential rotation through a solar cycle.}

\section{Global Convection}

The parametric regime in which global convection operates is extreme \citep[Prandtl number $\approx 10^{-6} - 10^{-4}$, Rayleigh number $\approx 10^{19} - 10^{24}$, and Reynolds 	number $\approx 10^{12} - 10^{16}$:][]{miesch05}, so 3D direct numerical simulations of the full solar convective envelope that can incorporate this vast range of scales are likely impossible for the foreseeable future. Similarly, it is very difficult to perform laboratory experiments of convection in strongly stratified fluids in these parameter ranges. Unlike for surface convection, it is not obvious whether there are easy-to-model and dominant physical processes that, when modeled, {would assure us of an accurate description of the large scales.}

\begin{figure}
   \centerline{
               \includegraphics[width=0.99\textwidth,clip=]{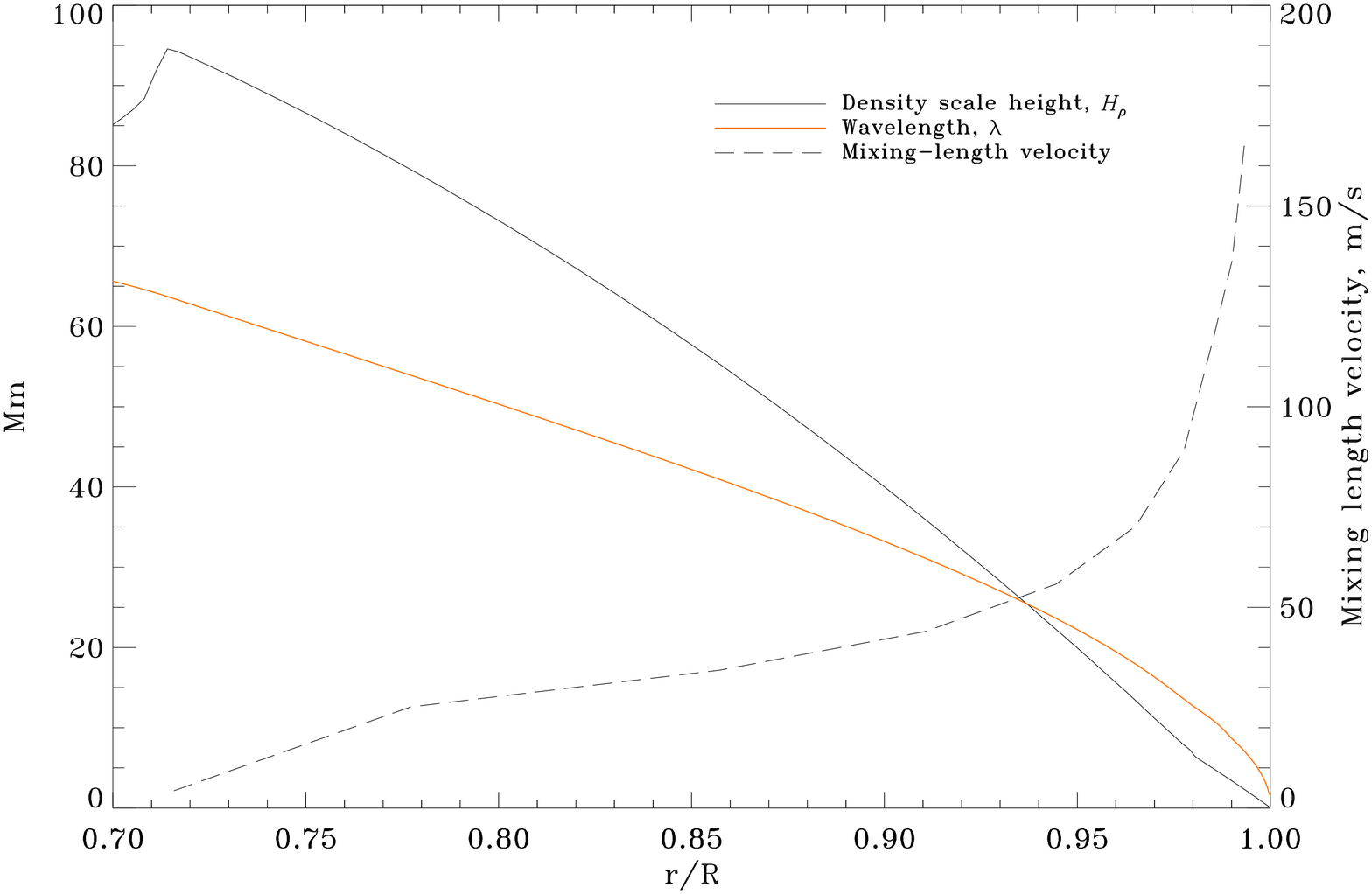}}\vspace{0.5cm}
   \caption{The mixing length, velocity scale, and acoustic wavelength plotted as a function of depth. Mixing-length theory \citep{spruit74}
   posits that convection is effected by parcels moving one length scale at one velocity. The mixing length is roughly the density scale height
   \citep[$H_\rho \approx 1.8 \times H_p$, where $H_p$ is the pressure scale height and 1.8 $H_p$ is the mixing length estimated
   from simulations; see, \eg][]{trampedach11}. Roughly speaking seismic measurements are diffraction limited, and the local acoustic wavelength
   is indicative of the cutoff scale for making inferences. Thus, if mixing length were applicable to the Sun, we would likely be able to
   image convective structures in the interior because they would be larger than the local acoustic wavelength.}\label{MLT}
  \vspace{-0.2cm}
  \end{figure}

In mixing-length phenomenology, thermal transport is effected by parcels of fluid of specified spatial and velocity scales, coherent over a length scale (termed the {\it mixing length}). Despite the simplicity of this model \citep[see, \eg][]{cox_2004}, it has been successful in predicting the depth of the convection zone and the dominant scale and velocity magnitude of surface convection \citep[with suitable additional assumptions, see][]{nordlund09}. The predicted scales and convective velocities are shown in Figure~\ref{MLT}.
	
Because density and pressure scale heights increase with depth, mixing-length theory indicates a corresponding increase in spatial convective scale (note also that increasing densities with depth imply enhanced heat capacities which, in turn, result in correspondingly lower velocity magnitudes). These large convective cells, the so-called {\it giant cells}, have also been widely reported in numerical simulations \citep[\eg][]{miesch_etal_08}. Invoking the {\it anelastic approximation} \citep{gough69}, a regime describing low-Mach-number and strongly non-Boussinesq stratified convection, 3D simulations of global convection have been performed at increasing resolutions over the past few years \citep{miesch_etal_08,charbonneau10,kapyla1, kapyla2}. A common feature of computation and mixing-length theory is that there is a great deal of overturning convection, resulting in substantial kinetic energies (not exceeding 15\,\% of the net energy flux; also a strong function of depth). It has been estimated that such vigorous overturning results in the outward transport of $O(10^4)$ solar luminosities and $O(10^4)- 1$ inwards, with a sum total of one solar luminosity outwards \citep[M. Sch\"{u}ssler 2012, private communication; also see][]{spruit97}. If this is indeed the regime that describes solar-interior convection, it certainly is not very efficient! There is, of course, no reason to think that non-equilibrium processes such as thermal convection obey any rules of optimizing efficiency. Indeed, turbulence literature is replete with examples to the contrary. For instance, while the energy production needed to maintain a turbulent wall flow is of the order unity, the production and dissipation near the wall are both of the order of the flow Reynolds number, one almost canceling the other, but diffusing outward the small difference between them. This difference is what maintains the flow.

\subsection{Measurements and Seismic Constraints}
{Considerable efforts have been devoted to attempting surface \citep{balle86, hathaway00} and interior \citep{duvall, duvall03} detection of giant convective cells, but the evidence supporting their existence is generally weak. It is thought that giant cells possess very low surface-velocity amplitudes due to screening by the subsurface shear layer \citep{miesch_etal_08}.}

\begin{figure}
   \centerline{
               \includegraphics[width=\textwidth,clip=]{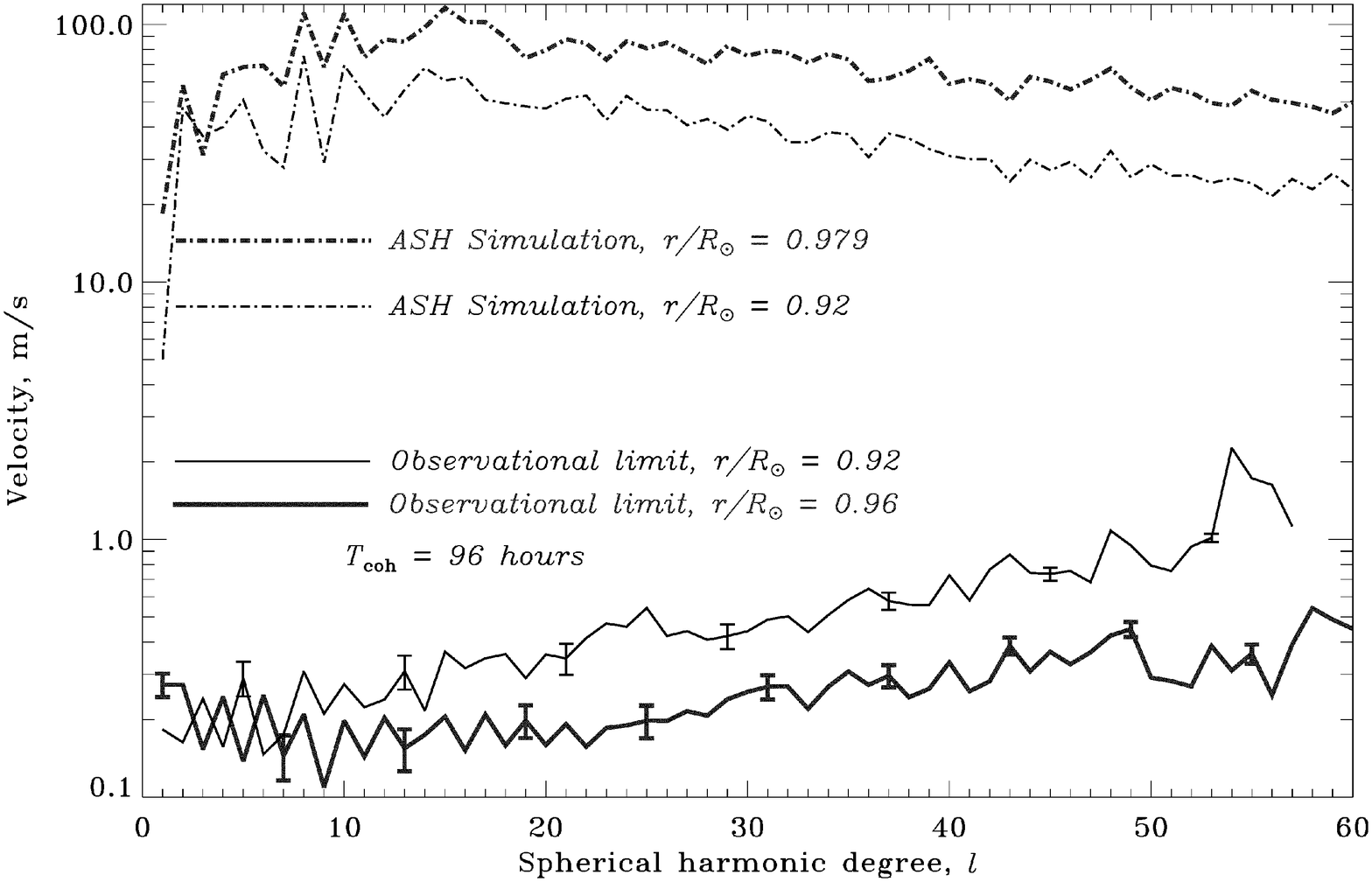}}\vspace{0.5cm}
   \caption{Seismic upper bounds \citep{hanasoge12_conv} on the amplitude of convective velocities in the solar interior at
   depths of $r/\rsun = 0.92, 0.96$. The bounds are compared with spectra derived at corresponding
   depths from the ASH simulations of \citet{miesch_etal_08}. In deriving the observational constraints, a coherence
   time of 96 hours for the convective cells, consistent with estimates from mixing-length theory \citep{spruit74} was assumed. Given that the photospheric-velocity spectrum in Figure~\ref{spectrum}
   is also weak at large scales of $\ell\lesssim 60$, it is not surprising that the convective-velocity amplitudes are
   so low at these depths.}\label{constraints}
  \vspace{-0.2cm}
  \end{figure}

Using techniques of time--distance helioseismology, \citet{hanasoge10} and \citet{hanasoge12_conv} placed stringent bounds on the interior convective-velocity spectrum. Two-point correlations measured from finite temporal segments of the observed line-of-sight photospheric Doppler velocities, taken by HMI \citep{hmi}, were used in the analysis. These correlations were spatially averaged according to a deep-focusing geometry \citep{hanasoge10} in order to image the interior. Convective coherence timescales \citep{spruit74, gough77, miesch_etal_08} were taken into account in choosing the temporal length of the data.
By construction, these measurements are sensitive to the three components of the underlying flow field, i.e. longitudinal, latitudinal, or radial, at specific depths of the solar interior ($r/\rsun = 0.92, 0.96$), shown in Figure~\ref{constraints}.

{Analyzing motions of supergranules, \citet{hathaway2013} claim to have discovered large-scale flow systems (which could be convection) at the photosphere, on the order of 8 ${\rm ms^{-1}}$ in magnitude.
Given such weak photospheric motions, and keeping in mind that the density is about $10^4$ times larger at a depth of $r = 0.96 \rsun$ than at the photosphere, it is likely that convective motions on large scales will become significantly weaker in magnitude at depth (unless there is a screening effect, in which case the velocity magnitude would increase with depth). The seismic constraints in Figure~\ref{constraints} could likely comfortably accommodate such weak motions. Thus the problem
of weak Reynolds stresses on large scales cannot be explained away by the claims of \citet{hathaway2013}. The basic issue is that
the photospheric power spectrum of convection drops linearly with $\ell$ at very large scales, and this has yet to be understood. }

This {large} difference between observations and simulations suggests that deep convection in the Sun may be operating under conditions differing from a mixing-length scenario. A plume-based thermal-transport mechanism in this alternative regime has been the subject of speculation by \citet{schmitt84}, \citet{rieutord95}, \citet{spruit97}
and was explored further by \citet{rempel2005}.
{\citet{rieutord95} and \citet{spruit97} envisaged a scenario in which very weak upflows are seeded at the base of the convection zone, buoyantly rising and undergoing expansion due to the decreasingly dense layers they encounter. These flows are in mass balance with cool intergranular descending plumes formed at the photosphere, which undergo compression as they fall into layers of increasing density in the interior. The descending plumes entrain fluid as they fall through the convection zone, but not enough to be completely mixed. This scenario can be likened to a cold sleet amid warm upwardly moving plasma. The associated turbulent kinetic energy in the convection zone would be very weak and this thermal transport process as a whole would elude detection because the upflows are too weak and the downflows too small in spatial dimension.}
This mechanism {would support} outward thermal transport of a solar luminosity's worth of heat flux at low kinetic energy. However, the transport of angular momentum, i.e. the maintenance of differential rotation and meridional circulation, is not as easily explained. Based on scaling arguments, \citet{miesch12} estimate the minimum convective kinetic energy (and associated Reynolds stresses) required to sustain these large-scale flow circulations. \citet{gizon12} provide an interesting comparison between seismology, simulation, and the phenomenology of \citet{miesch12}.

\section{Progress on Global Convection}
High-quality observations taken by HMI in tandem with improvements in the theory to interpret seismic measurements will allow us to make much better inferences of solar-interior magneto-convection. However, to truly make progress, {it is important to put forth} a set of predictions from theory and computation, along the
lines of \citet{kitchatinov05}, \citet{kapyla07}, and \citet{miesch12}. What are the minimal set of constraints on convective motions so that a solar luminosity's worth of heat flux can be transported outwards while differential rotation and meridional
circulation are sustained? (Additional constraints might also incorporate the tachocline and the global dynamo.) What seismic measurements would most clearly suggest one mechanism or theory over another and what theories can be eliminated using these measurements? It is the role of convection theorists to help guide seismology towards
making the most appropriate and useful measurements (when it is possible to make them).

The constraints derived by \citet{hanasoge12_conv} provide, at best, a starting point for a significant undertaking towards measuring convective velocity amplitudes in the Sun. \citet{hanasoge12_conv} suggest that their upper bounds are conservative and may, with better modeling, be able to drive down these bounds even further. Such a situation,
were it to be shown true, would challenge our current understanding of global dynamics.
To paraphrase \citet{spruit97}: ``{convection in the Sun is driven by cooling from the top, not heating from the bottom}". Indeed, the {largest-amplitude} fluid motions occur at the upper thermal boundary layer ($\approx 300$ km), which also happens to be much thinner than the lower convective boundary layer \citep{monteiro11}. The kinetic energy in the convection zone proper is tightly connected to the fate of the innumerable minuscule diving plumes formed at intergranular lanes. {Do these plumes disintegrate as in a variety of numerical simulations or do they continue on unaffected through the convection zone \citep[as][would have it]{spruit97}? These plumes likely drive turbulence by injecting energy at small scales \citep[\eg][]{rieutord10} and, by this chain of reasoning, the surface plays an asymmetrically important role in controlling the dynamics of the convection zone.}

\citet{balbus12} suggest that centrifugal distortion at the base of the convection zone due to solar rotation
sets up a latitudinal entropy gradient, which in turn drives differential rotation. However, as \citet{miesch12} point out, temperature gradients alone cannot set up and maintain the rotation, especially in the presence of meridional circulations (see also Sections~\ref{latgrad} and~\ref{stress}). {It may well be possible that the combination of Reynolds stresses associated with vigorous near-surface turbulence and a latitudinal temperature gradient created by centrifugal distortion at the base of the convection zone are the dual mechanisms that create the observed global fluid dynamical circulations in the Sun. Needless to say, the validity of this hypothesis is subject to further testing.}

\section{Speculative Concluding Remarks}

{In this review, we have concerned ourselves principally with two types of scales:  supergranules and those of deep convection.
The energy carrying scale, namely granulation is generally well understood \citep[\eg][]{nordlund09} and has therefore not been discussed much.
Surface granulation (a shell whose thickness is about 0.2\,\% of the Sun's radius) is likely not driven by deep convective motions, and the dynamics of granules are determined by a combination of strong stratification and equation-of-state considerations.}

With respect to deep convection, a variety of concepts ranging from mixing length to giant plumes to structures in between have been invoked. We have listed several of them and pointed out the potential role of helioseismology in addressing such questions if they are well-posed. In any case, since one cannot observe interior convection directly, we can at best hope for a self-consistent scenario that is also consistent with every known fact. Such a scenario has not yet emerged.
Supergranules present a different class of uncertainty. One can observe them (admittedly, statistically significant measurements require massaging of the data), but yet cannot answer fundamental questions about their physics. What remains clear, however, is that their thermal signature is weak, {which would not be the
case if it were a conventional convective process}.

The first question that one needs to answer is whether supergranules {are linked to surface granular flows} or arise as a surface manifestation of interior motions. The fact that their length and time scales are large compared to those of surface granules is not in itself an adequate reason to dismiss the possibility that they belong to the family of granules themselves. For example, in canonical turbulent flows (\eg~pipe flow), one can detect motions whose length scales are some 30\,--\,100 times larger than the largest structures that one may expect (\eg the pipe diameter). These superlong structures arise from a weak coalescence of many scales, each of the order of the diameter, probably coupled through the pressure field. The coupling may also simply be statistical in nature akin to those seen in percolation phenomenon. In the Sun, the fact that the outer shell is likely not of uniform thickness with unchanging fluid properties may itself generate motions other than standard granular structure. If one were able to simulate convection motions using an unsteady shell thickness and time-varying properties, one may go some way in answering the question just posed.

{If, on the other hand, it were the case that supergranules result from the interaction of deep convection structures, we should at least ask: how does the energy in convective scales on the order of supergranules compare to the energy of supergranules at various depths? The answer is unclear: neither simulations nor helioseismic measurements possess the full set of ingredients needed to answer this question. Helioseismology probes scales that are likely too coarse (at depth) and ASH simulations do not include surface layers.
We believe that other sensible questions follow once this basic issue is addressed. One obvious way to progress is to improve the resolution of numerical simulations (and reconcile with helioseismology results). Likewise, coming from the opposite end, it is important to improve the accuracy and resolution of helioseismic machinery.}

\acknowledgements
  SMH acknowledges funding from NASA grant NNX11AB63G and thanks the Courant Institute for its hospitality.

\bibliography{ms}\bibliographystyle{spr-mp-sola-cnd}
\end{article}

\end{document}